\documentclass[12pt,fleqn]{article}
\usepackage{amssymb}
\usepackage{amsmath}
\usepackage[dvips]{graphicx}
\begin{document}
%----------------------------------------------------------
%----------------------------------------------------------
%----------------------------------------------------------
%----------------------------------------------------------
\begin{flushright}
KEK-TH-908\\
{\tt hep-ph/0308021 }\\
\end{flushright}
%----------------------------------------------------------
%----------------------------------------------------------
%----------------------------------------------------------
%----------------------------------------------------------
\vspace*{1.5cm}
\begin{center}
    {\baselineskip 25pt
    \Large{\bf 
    %%%%%%%%%%%%%%%%%%%%%%%%%%%%%%%%%%%%
    %%%%%% TITLE %%%%%%%%%%%%%%%%%%%%%%%%%%
    %%%%%%%%%%%%%%%%%%%%%%%%%%%%%%%%%%%%

    Higgs-mediated muon-electron conversion process \\
    in supersymmetric seesaw model
    
    %%%%%%%%%%%%%%%%%%%%%%%%%%%%%%%%%%%%
    %%%%%%%%%%%%%%%%%%%%%%%%%%%%%%%%%%%%
    %%%%%%%%%%%%%%%%%%%%%%%%%%%%%%%%%%%%
    }
    }

\vspace{1.2cm}
\def\thefootnote{\fnsymbol{footnote}}
{ Ryuichiro Kitano,$^{a,}$}\footnote
{email: {\tt kitano@ias.edu}}
{ Masafumi Koike,$^{b,}$}\footnote
{email: {\tt mkoike@post.kek.jp}}
{ Shinji Komine,$^{b,}$}\footnote
{email: {\tt komine@post.kek.jp}}
and
{ Yasuhiro Okada$^{b,c,}$}\footnote
{email: {\tt yasuhiro.okada@kek.jp}}
\vspace{.5cm}
    
{\small {\it
$^a$School of Natural Sciences, Institute for Advanced Study,
Princeton, NJ 08540 \\
\vspace*{2mm}
$^b$Theory Group, KEK, Oho 1-1, Tsukuba, 
Ibaraki 305-0801, Japan \\
\vspace*{2mm}
$^c$Department of Particle and Nuclear Physics,
The Graduate University for Advanced Studies,\\
Oho 1-1, Tsukuba, Ibaraki 305-0801, Japan
}}
    
    \vspace{.5cm}
    \today
    
    \vspace{1.5cm}
    {\bf Abstract}

\end{center}

\bigskip
%%%%%%%%%%%%%%%%%%%%%%%%%%%%%%%%%%%%%%%
%           ABSTRACT
%%%%%%%%%%%%%%%%%%%%%%%%%%%%%%%%%%%%%%%

We study
the effect of the Higgs-exchange diagram for the lepton flavor violating 
muon-electron conversion process in nuclei
in the supersymmetric seesaw model.
The contribution is significant 
for a large value of $\tan \beta$ and a small value of a neutral heavy 
Higgs boson mass, in which case the ratio of the branching ratios of
$B(\mu N \to e N) / B(\mu \to e \gamma)$
is enhanced.
We also show that the target atom dependence
of the conversion branching ratio provides
information on the size of the Higgs exchange diagram.

%%%%%%%%%%%%%%%%%%%%%%%%%%%%%%%%%%%%%%%

\newpage
\def\thefootnote{\arabic{footnote}}
\setcounter{footnote}{0}
%\baselineskip 20pt
%\baselineskip 17.6pt

%------------------------------------------------------------
%                 START
%------------------------------------------------------------

Among various candidates for physics beyond the standard model, 
the supersymmetric (SUSY) extension is considered to be one of the most
promising ones, 
providing us with a solution to the hierarchy problem.
In addition, the gauge coupling constants measured precisely in the last 
decade show a remarkable agreement with the prediction of the SUSY
grand unified theory (GUT) \cite{susy-gut-coupling}.
Although direct searches for SUSY particles are the most important,
it is also interesting to see implications of such a
theory in the low-energy phenomena. This will be important even after we 
discover SUSY particles at energy frontier experiments, because some of 
model parameters can be only accessible from low energy experiments.

Lepton flavor violation (LFV) offers a possibility to explore 
SUSY models from low energy experiments \cite{Kuno:1999jp}.
In the standard model, the lepton number is conserved separately 
for each generation, so that LFV in charged lepton precesses is 
forbidden. On the other hand, recent developments in neutrino physics
indicate that the lepton flavor is not conserved in the neutrino sector.
However, the simplest model of a finite neutrino mass, namely the see-saw
model, does not induce observable effects of LFV in charged 
lepton processes since only the mass terms of the neutrinos violate the 
lepton flavor conservation
\cite{Bilenkii:1977du}.
The situation is completely different in the context of SUSY. In SUSY
models the LFV in muon and tau decays is considerably enhanced due to
the existence of the scalar partner of leptons, and therefore the
branching ratios of these processes can be close to the reaches of
current or near future experiments in SUSY-GUT \cite{susy-gut} as well 
as SUSY seesaw models
\cite{Borzumati:1986qx,Hisano:1996cp,Hisano:1995nq}.

In the SUSY model, a new source of LFV appears in the off-diagonal
components of the slepton mass matrices.  In the case of the SUSY
seesaw model, the Dirac Yukawa interactions of the neutrinos induce
the off-diagonal components at the one-loop level,
even if we assume the slepton mass matrix to be proportional to the
unit matrix at the high-energy scale such as in the minimal
supergravity scenario \cite{Borzumati:1986qx}.
This effect can be sizable, and the current upper bound of the
branching ratio of the $\mu \to e \gamma$ decay already puts severe
constraints on the SUSY parameters.

Another important process is the $\mu-e$ conversion in nuclei.
In the effective Lagrangian at the energy scale of the muon mass, this
process can be induced by several four-fermion operators, in addition
to the photon dipole-type operator, which is responsible for the $\mu
\to e \gamma$ decay.
If the latter is the only source of LFV, the branching ratio
for the $\mu-e$ conversion is suppressed roughly by $O(\alpha)$ 
compared to the branching ratio of the $\mu \to e \gamma$ decay.
Even if this is the case, significance on new physics search from
two processes can be similar since the experimental upper limit 
is lower for the $\mu-e$ conversion process.  The current best 
experimental upper bounds for the branching fractions are 
$B(\mu \to e \gamma) < 1.2 \times 10^{-11}$ 
\cite{Brooks:1999pu}
and 
$B(\mu {\rm Ti} \to e {\rm Ti}) < 6.1 \times 10^{-13} $
\cite{SINDRUM-II}, 
respectively. There are several planned
experiments which are aiming at improving the bounds 
of the branching fractions for relevant
processes by three or four orders of magnitudes
\cite{PSI,MECO,PRISM}. 
If other four-fermion interaction is sizable, or
even dominant, the $\mu-e$ conversion branching ratio
may not be suppressed by $O(\alpha)$ relative to the $\mu \to e \gamma$ 
branching ratio. 
For example, in R-parity violating SUSY models,
the contribution from the scalar type four-fermion 
interaction is shown to be important
especially through the strange-quark
\cite{Faessler:pn} and the bottom-quark couplings \cite{Kosmas:2001sx}.

Recently, 
the effect of the neutral Higgs exchange diagrams 
in the various flavor changing neutral current (FCNC) processes 
is considered, and a possibility of large contributions is pointed out
especially for large $\tan\beta$ and small $m_A$ region
in SUSY models
\cite{FCNC}.
In LFV processes,
the new effect to the $\tau \to 3 \mu$ and $\tau \to \mu \eta$ decay
is studied
\cite{Babu:2002et, Dedes:2002rh, Sher:2002ew}.
Since the Higgs-mediated FCNC does not contribute 
to the $\tau \to \mu \gamma$ decay,
the ratio of $B(\tau \to 3 \mu) / B(\tau \to \mu \gamma)$
is useful to reveal the existence of the effect.

In this letter,
we studied the effect of the Higgs-exchange diagram
on the $\mu - e$ conversion process
in the SUSY seesaw model.
In contrast to the $\mu \to 3e$ decay,
the Higgs-mediated contribution to
the $\mu - e$ conversion process is not suppressed by
the electron mass but only by the nucleon masses,
because the Higgs-boson coupling to the nucleon is
shown to be  characterized by the nucleon mass using the conformal 
anomaly relation
\cite{Shifman:zn}.
The most important contribution turns out to come from 
the exchange of the heavier scalar Higgs boson $(H^0)$ which couples 
to the strange quark scalar current in the nucleus.
We found that the transition amplitude 
from this type of diagrams becomes fairly large
compared to the photon-exchange diagram responsible for
the $\mu \to e \gamma$  decay 
in the large $\tan \beta$ and the light $H^0$ region.
Therefore,
the ratio of $B(\mu N \to e N) / B(\mu \to e \gamma)$
is quite sensitive to the Higgs-exchange effect, just as 
$B(\tau \to 3 \mu) / B(\tau \to \mu \gamma)$
is important in $\tau$ decays.
Also, we show that 
it is possible to identify  
the Higgs-mediated LFV effect
by looking at the target atom dependence of the
branching ratio, e.g.\
$B(\mu {\rm Pb} \to e {\rm Pb}) / B(\mu {\rm Al} \to e {\rm Al})$.

In the SUSY seesaw model, the off-diagonal components of the slepton mass
matrix appear in the left-handed sleptons through the neutrino Yukawa
interactions.
We assume that the slepton mass matrix is proportional to the unit matrix
at the GUT scale, and evaluate  the effects of the neutrino Yukawa interaction
to the slepton sector.
The superpotential of the lepton sector is given by
$W = f_e^i H_1 \cdot E^c_i L_i
+ f_\nu^{ij} H_2 \cdot N^c_i L_j 
+ (1/2) M_N^{i} N_i^c N_i^c$, 
where $H_1$ and $H_2$ are the doublet Higgs fields,
$L_i$, $E^c_i$, and $N^c_i$ are the superfields corresponding to 
the left-handed leptons,
right-handed leptons, and right-handed neutrinos of the $i$-th
generation, respectively.
The neutrino mass matrix is obtained by integrating out the heavy
right-handed neutrinos as $ m_{\nu}^{ij} = (f_\nu^T
M_N^{-1} f_\nu )_{ij} v^2 \sin^2 \beta/2$ \cite{seesaw}, where $v$ is the vacuum
expectation value (VEV) of the Higgs field ($v = 246$ GeV) and the
angle $\beta$ is defined by $\tan \beta = \langle H_2^0 \rangle /
\langle H_1^0 \rangle$.
To obtain the correct size of the neutrino masses,
the right-handed neutrinos should be as heavy as $10^{14}$ GeV
for $f_\nu \sim O(1)$.
The Yukawa interactions represented by $f_\nu$ violate the lepton
flavor conservation. This violation is imprinted to the slepton mass
matrix in the low-energy Lagrangian.
The renormalization group equation (RGE) 
running effect induces the off-diagonal components
in the left-handed slepton mass matrix which are approximately
given as follows:
\begin{eqnarray}
 (\Delta m_{\tilde{l}_{L}}^2)_{ij}
\simeq - \frac{1}{8 \pi}
f_\nu^{ki*} f_\nu^{kj} m_0^2 (3 + |a_0|^2) \log \frac{M_{\rm GUT}}{M_N^k}
\ ,
\label{app-delta-msq}
\end{eqnarray}
where the SUSY breaking parameters $m_0$ and $a_0$ represent
scalar masses and three point scalar interactions 
at the GUT scale, respectively.

LFV in the Higgs coupling originates from 
the non-holomorphic correction to the Yukawa
interactions of the charged leptons
\cite{Babu:2002et}.
One-loop diagrams mediated by sleptons
induce the following 
Yukawa interaction terms:
\begin{eqnarray}
 {\cal L} = f_e^{i} \bar{e}_i P_L e_i H_1^0
+ f_e^{i} \bar{e}_i 
\left(
\epsilon_1^{(i)} \delta_{ij} + \tilde{\epsilon}_2^{(ij)}
\right)
P_L e_j H_2^{0*} + {\rm h.c.}\ ,
\label{Yukawa}
\end{eqnarray}
where $P_{L,R} = (1 \mp \gamma_5)/2$.
The non-holomorphic interactions
$\epsilon_1$ and
$\tilde{\epsilon}_2$ are given by\footnote{
There are sign differences in theses equations compared to the results 
in Refs.~\cite{Babu:2002et,Dedes:2002rh}.
}
\begin{eqnarray}
 \epsilon_1^{(i)} &=& 
g_Y^2 \mu M_1
\left[
I_3(M_1^2, m_{\tilde{e}_{Ri}}^2, m_{\tilde{l}_{Li}}^2)
+
\frac{1}{2} I_3 (M_1^2, \mu^2, m_{\tilde{l}_{Li}}^2)
- 
I_3 (M_1^2, \mu^2, m_{\tilde{l}_{Ri}}^2)
\right]
 \nonumber \\
&& 
-
\frac{3}{2} g_2^2 \mu M_2
I_3(M_2^2, \mu^2, m_{\tilde{l}_{Li}}^2)
\ ,
\end{eqnarray}
\begin{eqnarray}
 \tilde{\epsilon}_2^{(ij)} &=& 
- g_Y^2 \mu M_1 (\Delta m_{\tilde{l}_{L}}^2)_{ij}
\left[
I_4 (M_1^2, m_{\tilde{e}_{Ri}}^2, m_{\tilde{l}_{Li}}^2, m_{\tilde{l}_{Lj}}^2)
+ \frac{1}{2} 
I_4 (M_1^2, \mu^2, m_{\tilde{l}_{Li}}^2, m_{\tilde{l}_{Lj}}^2)
\right]
\nonumber \\
&& + \frac{3}{2}
g_2^2 \mu M_2 (\Delta m_{\tilde{l}_{L}}^2)_{ij}
I_4 (M_2^2, \mu^2, m_{\tilde{l}_{Li}}^2, m_{\tilde{l}_{Lj}}^2)\ ,
\end{eqnarray}
where $g_Y$ and $g_2$ are the gauge coupling constants, and
$M_1$, $M_2$, and $\mu$ are the gaugino and Higgsino mass parameters.
The above formulas are based on the calculation of the effective 
Yukawa interaction in the SU(2)$_L$ $\times$ U(1)$_Y$ symmetric limit.  
The mass parameters $m_{\tilde{e}_{Ri}}^2$ and $m_{\tilde{e}_{Li}}^2$
are the slepton masses for the $i$-th generation. 
The functions $I_3$ and $I_4$ are defined by
\begin{eqnarray}
 I_3 (a, b, c) 
= - \frac{1}{(4 \pi)^2}
\frac{ab \log (a/b) + bc \log (b/c) + ca \log (c/a)}
{(a-b)(b-c)(c-a)}\ ,
\end{eqnarray}
\begin{eqnarray}
 I_4 (a,b,c,d)
= \frac{1}{(4 \pi)^2}
\!\!\! && \!\!\! \left[ \ 
\frac{a \log a}{(b-a)(c-a)(d-a)}
+
\frac{b \log b}{(a-b)(c-b)(d-b)}
\right.
\nonumber \\
&&
\left.
+ \frac{c \log c}{(a-c)(b-c)(d-c)}
+
\frac{d \log d}{(a-d)(b-d)(c-d)}\ \ 
\right]\ .
\end{eqnarray}
Note that
the parameters $\epsilon_1$ and $\tilde{\epsilon_2}$
do not vanish even in the limit of large masses of SUSY particles.
This is quite different from the photon-exchange diagrams of LFV, where
the amplitude becomes small for large masses of internal SUSY particles.

The Yukawa interactions in Eq.(\ref{Yukawa})
can be written in terms of the fields in the mass eigenstates.
For the $\mu - e$ transition,
the Lagrangian is given by
\begin{eqnarray}
 {\cal L} =
- \frac{m_\mu \kappa_{21}}{v \cos^2 \beta}
( \bar{\mu} P_L e ) 
\left[
\cos (\alpha - \beta ) h^0
+ \sin (\alpha - \beta ) H^0
- i A^0
\right] + {\rm h.c.}\ ,
\label{Yukawa2}
\end{eqnarray}
where $h^0$ and $H^0$ are the scalar Higgs fields ($m_{h^0} < m_{H^0}$),
and $A^0$ is the pseudoscalar Higgs field.
The LFV parameter $\kappa_{21}$ is given by
$ \kappa_{21} =
\tilde{\epsilon}_2^{(21)} / ( 1 + \epsilon_1^{(2)} \tan \beta )^2$.
In the limit where the masses of $H^0$ and $A^0$ go to infinity,
the LFV interaction of the lightest Higgs boson vanishes
since the standard model does not have LFV.
Therefore the contributions from $H^0$ and $A^0$ are
important in LFV processes  for relatively small values of heavy 
Higgs boson masses.

The $\mu - e$ conversion process occurs through this interaction
by the exchange of the Higgs boson with a nucleus.
The effective four-fermion interactions are given by
\begin{eqnarray}
 {\cal L}_{\rm eff} =
- \frac{ m_\mu \kappa_{21} }{v \cos^2 \beta}
\sum_{q = u,c,t}
\left[ {\rule[-3mm]{0mm}{10mm}\ } \right. \!\!\!
\!\!\!&&\!\!\!
\frac{m_q}{v \sin \beta} 
\left \{ 
- \frac{\cos (\alpha - \beta) \cos \alpha}{m_{h^0}^2}
- \frac{\sin (\alpha - \beta) \sin \alpha}{m_{H^0}^2}
\right \} ( \bar{e} P_R \mu )( \bar{q} q )
\nonumber \\
\!\!\!&&\!\!\!
- \frac{m_q}{v \sin \beta} 
\frac{\cos \beta}{m_{A^0}^2} (\bar{e} P_R \mu )(\bar{q} \gamma_5 q)
\left. {\rule[-3mm]{0mm}{10mm}\ } \right]
\nonumber \\
- \frac{ m_\mu \kappa_{21} }{v \cos^2 \beta}
\sum_{q = d,s,b}
\left[ {\rule[-3mm]{0mm}{10mm}\ } \right. \!\!\!
\!\!\!&&\!\!\!
\frac{m_q}{v \cos \beta} 
\left \{ 
\frac{\cos (\alpha - \beta) \sin \alpha}{m_{h^0}^2}
- \frac{\sin (\alpha - \beta) \cos \alpha}{m_{H^0}^2}
\right \} ( \bar{e} P_R \mu )( \bar{q} q )
\nonumber \\
\!\!\!&&\!\!\!
- \frac{m_q}{v \cos \beta} 
\frac{\sin \beta}{m_{A^0}^2} (\bar{e} P_R \mu )(\bar{q} \gamma_5 q)
\left. {\rule[-3mm]{0mm}{10mm}\ } \right]
\ .
\label{L-eff}
\end{eqnarray}
The transition amplitude can be obtained by taking a matrix element.
We evaluate the amplitude of the coherent conversion processes
where the initial and final nuclei are in the ground state.
Compared to incoherent transition processes, the coherent processes are
expected to be enhanced by a factor of $O(Z)$ where $Z$ is 
the atomic number.
In those processes,
the matrix elements for the quark operators
are obtained in the following way
\cite{Kosmas:2001mv}.
The first step is to write down the effective Lagrangian 
in the nucleon level which is given by replacements of
\begin{equation}
\bar{q} q \to G_S^{(q,p)} \bar{p} p + G_S^{(q,n)} \bar{n} n
\quad
\mathrm{and}
\quad
\bar{q} \gamma_5 q \to G_P^{(q,p)} \bar{p} \gamma_5 p 
+ G_P^{(q,n)} \bar{n} \gamma_5 n \ , 
\label{replace}
\end{equation}
where $G$'s are coefficients which can be evaluated 
by taking matrix elements of quark operators by nucleon states.
Then we can take the matrix elements by a specific nucleus.
Since the initial and final states are the same,
the elements $\langle N | \bar{p} p | N \rangle$
and $\langle N | \bar{n} n | N \rangle$ are nothing but
the proton and the neutron densities in a nucleus in the non-relativistic
limit of nucleons.
In this limit, the other matrix elements
$\langle N | \bar{p} \gamma_5 p | N \rangle$ and 
$\langle N | \bar{n} \gamma_5 n | N \rangle$ vanish.
Therefore in the coherent $\mu-e$ conversion process,
the dominant contribution comes from the exchange of
$H^0$, not $A^{0}$.

As we can see in Eq.(\ref{L-eff}),
the LFV interactions contain factors 
of the muon and the quark masses
because we need two chirality flips
to form the scalar four-fermion operators.
However, there can be also  enhancement factors.
One is the $\tan \beta$ enhancement in the 
$H^0$ vertices,
with which 
the amplitude is proportional to $\tan^3 \beta$.
The other factor of  enhancement is due to the couplings of
the Higgs boson to the nucleons represented by
the $G_S^{(q,p)}$ and $G_S^{(q,n)}$ factors.
It is important to notice that the Higgs boson coupling to the 
nucleon is not suppressed by the up or down current quark masses
because it can strongly couple to the
gluons in the nucleon through the loop diagrams
of the quarks
\cite{Shifman:zn}.
Among the various quarks, 
the strange quark gives the dominant contribution
in the large $\tan \beta$ region,
since the down-type quarks have $\tan \beta$ enhancement
in the Yukawa coupling constants.
The values of the combinations of $m_s G_S^{(s,p)}$ and $m_s G_S^{(s,n)}$ 
turn out to be much larger compared to the contribution from the 
down quark and that from bottom quark diagrams.
The values are estimated to be
$m_d G_S^{(d,p)} / m_p = 0.029$, $ m_d G_S^{(d,n)} / m_n= 0.037$,
$m_s G_S^{(s,p)} / m_p  = m_s G_S^{(s,n)} / m_n = 0.21$, and
$m_b G_S^{(b,p)} / m_p = m_b G_S^{(b,n)} / m_n = 0.055$
\cite{Corsetti:2000yq}.

Once we write down the effective Lagrangian
in the nucleon level,
the estimation of the coherent conversion rate is straightforward
\cite{WF,Kitano:2002mt}.
The general interaction Lagrangian for the coherent conversion process
is given by
\begin{eqnarray}
 {\cal L}_{\rm int}
&=&
- \frac{4 G_{\rm F}}{\sqrt{2}}
\left(
m_\mu A_R \bar{\mu} \sigma^{\mu \nu} P_L e F_{\mu \nu}
+
m_\mu A_L \bar{\mu} \sigma^{\mu \nu} P_R e F_{\mu \nu}
+ {\rm h.c.}
\right)
\nonumber \\
&& -\frac{G_{\rm F}}{\sqrt{2}}
\sum_{\psi=p,n}
\left[ {\rule[-3mm]{0mm}{10mm}\ } \right. \!\!\!
\left(
\tilde{g}_{LS}^{(\psi)} \bar{e} P_R \mu 
+
\tilde{g}_{RS}^{(\psi)} \bar{e} P_L \mu 
\right) \bar{\psi} \psi
\nonumber \\
&& \hspace*{2.1cm}
+
\left(
\tilde{g}_{LV}^{(\psi)} \bar{e} \gamma^\mu P_L \mu 
+
\tilde{g}_{RV}^{(\psi)} \bar{e} \gamma^\mu P_R \mu 
\right) \bar{\psi} \gamma_{\mu} \psi
+ {\rm h.c.}
 \!\!\! \left. {\rule[-3mm]{0mm}{10mm}\ } \right]
\ ,
\label{general}
\end{eqnarray}
where $A$'s and $\tilde{g}$'s are dimensionless coupling constants.
The first two are the dipole operators which are
the same operators for the $\mu \to e \gamma$ decay.
We also have scalar and vector type four-fermion operators.
In the SUSY seesaw model,
since the slepton mixing appears only in the left-handed sleptons,
the parity is maximally violated, i.e.\ 
$A_L$, $\tilde{g}_{RS}^{(p,n)}$, $\tilde{g}_{RV}^{(p,n)}$
$\ll$
$A_R$, $\tilde{g}_{LS}^{(p,n)}$, $\tilde{g}_{LV}^{(p,n)}$.
All those operators are given through the one-loop diagrams
mediated by the SUSY particles
\cite{Hisano:1996cp}.
Besides the scalar operator from the Higgs exchange,
only the dipole operator is important
in the large $\tan \beta$ region
because of the dependence of $A_R \propto \tan \beta$.

With the above coefficients in the effective Lagrangian,
the conversion rate is simply given by
\begin{eqnarray}
 \omega_{\rm conv}
= 2 G_{\rm F}^2 
\left|
A_R^* D 
+ \tilde{g}_{LS}^{(p)} S^{(p)}
+ \tilde{g}_{LS}^{(n)} S^{(n)}
+ \tilde{g}_{LV}^{(p)} V^{(p)}
+ \tilde{g}_{LV}^{(n)} V^{(n)}
\right|^2
+ (L \leftrightarrow R)\ ,
\label{rate}
\end{eqnarray}
where $D$, $S^{(p)}$, $S^{(n)}$, $V^{(p)}$, and $V^{(n)}$
are the overlap integrals among
wave functions of the muon and the electron
and the nucleon densities or the electric field
in the nuclei.
For example, in the aluminum nuclei,
they are estimated to be
$D=0.0362 m_\mu^{5/2}$,
$S^{(p)} = 0.0155 m_\mu^{5/2}$,
$S^{(n)} = 0.0161 m_\mu^{5/2}$,
$V^{(p)} = 0.0167 m_\mu^{5/2}$, and
$V^{(n)} = 0.0173 m_\mu^{5/2}$
\cite{Kitano:2002mt}.

By comparing Eq.(\ref{L-eff}) and Eq.(\ref{replace})
with Eq.(\ref{general}),
we obtain the coupling constants 
$\tilde{g}_{LS}^{(p)}$ and $\tilde{g}_{LS}^{(n)}$,
and then we can derive the
formula for the 
the conversion rate by Eq.(\ref{rate}).
For example, in the aluminum and lead targets,
the conversion branching ratios
at large $\tan \beta$ are approximately given by
\begin{eqnarray}
B(\mu {\rm Al} \to e {\rm Al}) \simeq
 1.8 \times 10^{-4} \cdot
\frac{m_\mu^7 m_p^2 \kappa_{21}^2}{v^4 m_{H^0}^4 \omega_{\rm capt}} 
\tan^6 \beta\ ,
\label{app-Al}
\end{eqnarray}
and
\begin{eqnarray}
B(\mu {\rm Pb} \to e {\rm Pb}) \simeq
 2.5 \times 10^{-3} \cdot
\frac{m_\mu^7 m_p^2 \kappa_{21}^2}{v^4 m_{H^0}^4 \omega_{\rm capt}} 
\tan^6 \beta\ ,
\end{eqnarray}
respectively, where $\omega_{\rm capt}$ is the muon capture rate in
the nuclei.  The values are $\omega_{\rm capt} = 0.7054 \times 10^6
\mathrm{s}^{-1}$ and $13.45 \times 10^6 \mathrm{s}^{-1}$ in the
aluminum and the lead nuclei, respectively \cite{Suzuki:1987jf}.
If we take all the right-handed neutrino masses to be $10^{14}$ GeV,
the contribution of the Higgs exchange in Eq.(\ref{app-Al})
is roughly given by 
\begin{eqnarray}
 B(\mu {\rm Al} \to e {\rm Al})_{H^0} \sim O(10^{-13}) \cdot 
\left( \frac{200 {\rm GeV}}{m_{H^0}} \right)^4 \cdot
\left( \frac{\tan \beta}{60} \right)^6
\ ,
\end{eqnarray}
whereas
the contribution from the photon exchange is calculated to be
\begin{eqnarray}
B(\mu {\rm Al} \to e {\rm Al})_\gamma \sim 
O(10^{-13}) \cdot \left( \frac{1000 {\rm GeV}}{M_S} \right)^4
\cdot 
\left(
\frac{\tan\beta}{60}
\right)^2\ ,
\end{eqnarray}
where $M_S$ is defined by $M_S \equiv m_0 = M_{1/2}$ 
with the universal scalar mass $m_0$ 
and the gaugino mass $M_{1/2}$ at the GUT scale.
The above estimation shows that
the Higgs exchange is important for $\tan \beta \gtrsim 60$
and $m_{H^0} \ll M_S$ region
because of the different decoupling behavior.

We numerically calculate the branching ratios in order to discuss the
effect of Higgs boson exchange more quantitatively.
We take the universal 
soft masses for the squarks and sleptons $m_0$,
the gaugino mass $M_{1/2}$,
and the $A$-terms $a_0$ at the GUT scale.
In order to realize a relatively light Higgs mass spectrum, we take  
the quadratic mass parameters of the Higgs potential
$m^{2}_{H1}$ and $m^{2}_{H2}$ to
be independent. The values of the $\mu$ and $B$ parameters in
low energy are determined to reproduce correct vacuum expectation values.
Between the GUT and the right-handed neutrino mass scales,
the parameters evolve with 
the RGE of minimal SUSY standard model (MSSM) with the
right-handed neutrinos,
that induces $(\Delta m_{\tilde{l}_L}^2)_{ij}$ which is approximately
given in Eq.(\ref{app-delta-msq}).
The evolution of the parameters obeys the MSSM running below the right-handed
neutrino scale.
In the actual calculation we solve relevant RGE's numerically. 
Once we obtain the low-energy parameters,
we calculate the coefficients $A$'s and $\tilde{g}$'s
given by the one-loop diagrams
\cite{Hisano:1996cp}
and the Higgs exchange effects in Eq.(\ref{L-eff}).
Then we used the the values in Ref.~\cite{Kitano:2002mt}
to calculate the conversion branching ratios from the coefficients.

\begin{figure}[t]
\begin{center}
\includegraphics[width=14cm]{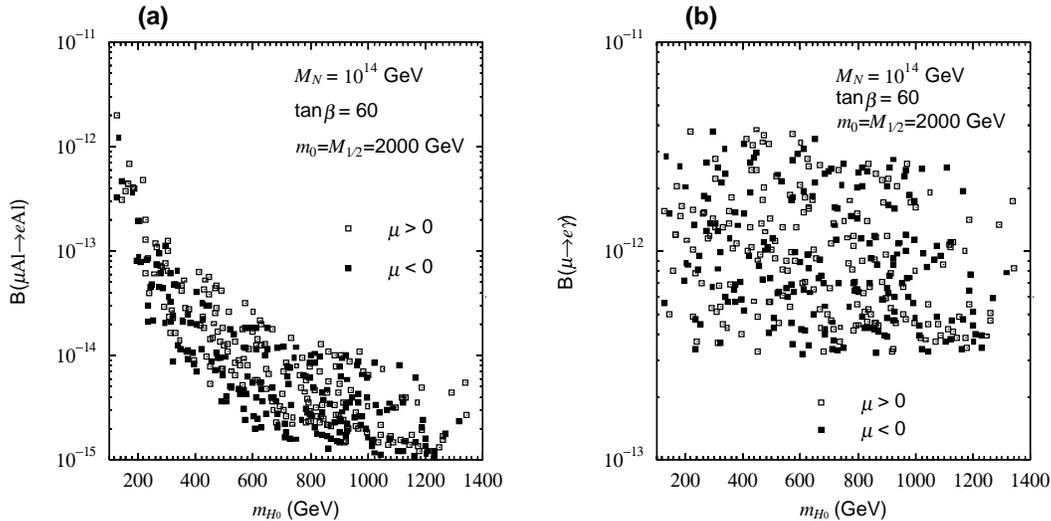} 
\end{center}
\caption{
The $m_{H^0}$ dependences of the branching ratios of the following
processes are shown: (a) $\mu-e$ conversion in aluminum nucleus and (b)
$\mu \to e \gamma$ decay.
We take the right-handed neutrino masses to be $10^{14}$ GeV, and $\tan
\beta = 60$.  The soft masses for the Higgs fields are treated as free
parameters.  } 
\label{fig1}
\end{figure}

\begin{figure}[t]
\begin{center}
\includegraphics[width=14cm]{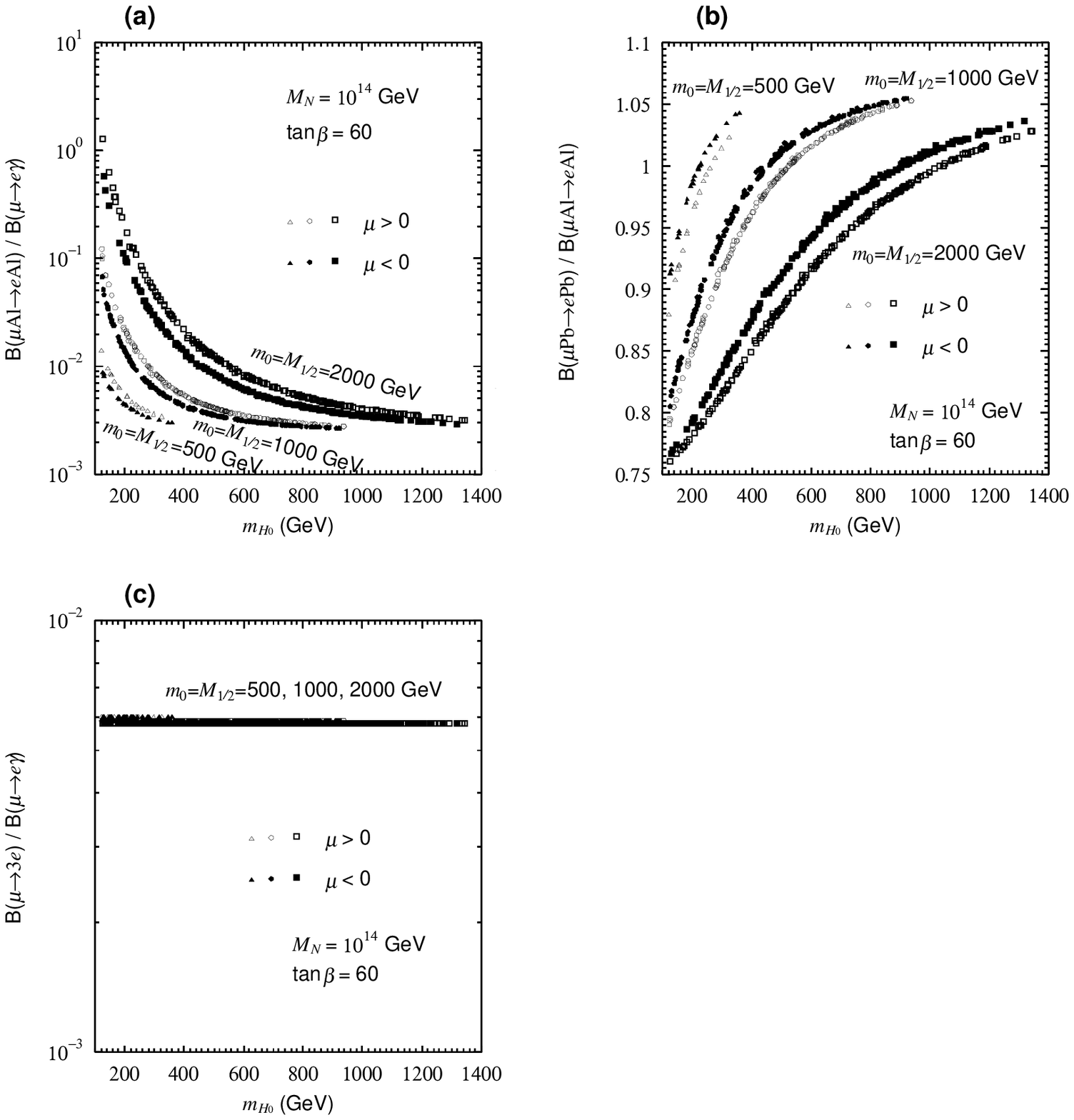} 
\end{center}
\caption{
The following ratios of the branching ratios are shown as functions of
$m_{H^0}$: (a) $B(\mu {\rm Al} \to e {\rm Al}) / B(\mu \to e \gamma)$,
(b) $B(\mu {\rm Pb} \to e {\rm Pb}) / B(\mu {\rm Al} \to e {\rm Al})$,
and (c) $B(\mu \to 3e) / B(\mu \to e \gamma)$.
We take the right-handed neutrino masses to be $10^{14}$ GeV, and $\tan
\beta = 60$.  The soft masses for the Higgs fields are treated as free
parameters.  } 
\label{fig2}
\end{figure}

The results of the calculation are shown in Figs.\ref{fig1} and \ref{fig2}.
We show in Fig.\ref{fig1}(a) the scatter plot of the value of $B(\mu
{\rm Al} \to e {\rm Al})$.
We fixed the $m_0$ and $M_{1/2}$ parameters to be 2000 GeV,
and $\tan \beta = 60$.
The right-handed neutrino masses are also fixed to
be $M_N^i = 10^{14}$ GeV for all the generations,
and the Yukawa coupling constants $f_\nu$ are determined 
so as to fit the neutrino oscillation data
of $\Delta m_{\rm atm}^2 = 3 \times 10^{-3}$ eV$^2$,
$\Delta m_{\rm sol}^2 = 4 \times 10^{-5}$ eV$^2$,
$\sin^2 \theta_{\rm atm} = 0.5$, and
$\sin^2 \theta_{\rm sol} = 0.25$
\cite{Fukuda:2001nk,Fukuda:1998mi,Ahn:2002up}.
We took the normal hierarchy with $m_{\nu 1} = 0$
and $U_{e3}=0$, where $U$ is the
Maki-Nakagawa-Sakata matrix
\cite{Maki:1962mu}.
As expected,
we can see the enhancement 
in the light $H^0$ region ($m_{H^0} \lesssim 600$ GeV)
because of the Higgs-exchange effect.
We can obtain the branching ratios 
for different values of $M_N$ by the scaling of
$B(\mu {\rm Al} \to e {\rm Al}) \propto (M_N)^{2}$.
Fig.\ref{fig1}(b) is the same plot for $B(\mu \to e \gamma)$.  The
enhancement in the light $H^0$ region is absent in the $\mu \to e
\gamma$ decay.
Therefore, taking the ratio of the branching ratios 
($B( \mu {\rm Al} \to e {\rm Al}) / B( \mu \to e \gamma)$),
we would see whether the contribution of the Higgs boson exchange is 
large.
Fig.\ref{fig2}(a) shows this ratio for 
the cases of $m_0 = M_{1/2} = 500$, $1000$, and
$2000$ GeV.
The ratio becomes large and can reach $O(1)$ for small $m_{H^{0}}$.
As increasing $m_{H^0}$,
it monotonically approaches to 0.0026,
which is the value predicted for the case that
the $\mu - e$ conversion occurs through 
photon exchange diagrams
\cite{Kitano:2002mt}.
The deviation from 0.0026 indicates that the existence of
the operators besides the photonic dipole operator.
Although this is an interesting prediction of the Higgs-exchange LFV,
it is not clear whether the Higgs-exchange effect is responsible for
the deviation when it is measured.

The target atom  dependence of the conversion branching ratio
is another interesting quantity which can discriminate
the different operators of the $\mu - e$ conversion process
\cite{Kitano:2002mt}.
There are three types of operators: 
dipole, scalar four-fermion, and vector four-fermion.
For light nuclei, all those operators show similar $Z$ dependences
in the $\mu - e$ conversion amplitude. Namely, 
the overlap integrals behave like
i.e.\ 
$D/(8e) = S^{(p)} = V^{(p)} \simeq ((A-Z)/A) S^{(n)} = ((A-Z)/A) V^{(n)}$,
in the non-relativistic limit of the muon wave function.
On the other hand, these overlap integrals take different values 
in the heavy nuclei due to 
the large relativistic effect.
For examples,
$B(\mu {\rm Pb} \to e {\rm Pb}) / B(\mu {\rm Al} \to e {\rm Al})$
for the dipole, the scalar, and the vector operators
are 1.1, 0.70, and 1.4, respectively
\cite{Kitano:2002mt}.
Thus the ratio of the branching ratio
in a heavy nucleus to that in light one provides information
on the type of operators responsible for the $\mu - e$ conversion.
In the case where the Higgs-exchange is dominated
as in the light $H^0$ region,
the heavy to light ratio corresponds to a value of the scalar-operator
prediction, which would be a robust indication of the effect.
We plot the ratio of 
$B(\mu {\rm Pb} \to e {\rm Pb}) / B(\mu {\rm Al} \to e {\rm Al})$
in Fig.\ref{fig2}(b).
We can see that the values indeed approach to the 
prediction of the scalar operator (0.70) in the light $H^0$ region,
and
they increase and approach asymptotically to the prediction
of the dipole operator (1.1)
as increasing $m_{H^0}$.
Measurement of this dependence can provide us with useful information
which will eventually lead us to extract the size of 
the Higgs-mediated LFV.

The enhancement due to the Higgs boson exchange is not as significant
in $\mu \to 3 e$ process as in $\mu - e$ conversion due to the small
Yukawa coupling of an electron.  The ratio $B(\mu \to 3 e) / B(\mu \to
e \gamma)$ is shown in Fig.\ref{fig2}(c).  There we can see the ratio
is almost constant (0.006) over the same parameter region as in
Fig.\ref{fig2}(a).

In summary,
we calculated the branching ratios of
the coherent $\mu - e$ conversion process
in the SUSY seesaw model.
The effect of the Higgs-mediated LFV
may give dominant contribution
in the large $\tan \beta$ and the light $H^0$ region
because of the $\tan^6 \beta$ and $(m_{H^0})^{-4}$
enhancement of the branching ratio.
The Higgs-exchange effect gives interesting signals 
on ratios of the branching ratios.
For example,
the ratio of $B(\mu {\rm Al} \to e {\rm Al}) / B(\mu \to e \gamma)$
is predicted to be 0.0026 with only the dipole operator, but
the Higgs-mediated process enhances this  ratio significantly.
Moreover, information on the operator responsible for the $\mu - e$
conversion can be obtained by the target atom dependence of the $\mu -
e$ conversion branching ratio.  This is particularly useful because we
can make a definite theoretical prediction for the ratio like $B(\mu
{\rm Pb} \to e {\rm Pb}) / B(\mu {\rm Al} \to e {\rm Al})$ depending
on the dominant type of operators.  For the ratio of $B(\mu {\rm Al}
\to e {\rm Al}) / B(\mu \to e \gamma)$, such definitive prediction is
possible only in the case of the photon-dipole-operator dominance.

\section*{Acknowledgments}

The work of R.K.\ was supported by DOE grant
DE-FG02-90ER40542.
The work of M.K.\ and S.K.\ was supported by the JSPS Research Fellowships
for Young Scientists.
The work of Y.O.\ was supported in part by a Grant-in-Aid of the 
Ministry of Education, Culture, Sports, Science and Technology, 
Government of Japan, No.~13640309 and No.~13135225.
%

%\newpage

\end{document}